\documentclass[12pt]{article}

\title{WHAT ARE QUANTUM FLUCTUATIONS?
ROUND TABLE OF THE THIRD CONFERENCE ON QUANTUM THEORY: RECONSIDERATION OF FOUNDATIONS}
\author{Andrei Khrennikov and Guillaume Adenier\footnote{
International Center for Mathematical
Modeling in Physics and Cognitive Sciences,
MSI, University of V\"axj\"o, S-35195, Sweden
Email:Andrei.Khrennikov@msi.vxu.se}\\ 
Theo M. Nieuwenhuizen\footnote{Department of Physics,  University of Amsterdam, The Netherlands}}

\date{}
\begin{document}

\maketitle

\begin{abstract} This is a transcript of the round table that took place during the 
conference 'Quantum Theory: Reconsideration of Foundations - 3', June 2005, V\"axj\"o, Sweden. 
There are presented opinions of leading experts in quantum foundations on such
 fundamental problems as the origin of quantum fluctuations and completeness of quantum mechanics.
\end{abstract}

\section{WHAT ARE QUANTUM FLUCTUATIONS?}

Theo M. Nieuwenhuizen: From the point of view of Stochastic Electro-Dynamics (SED), the world is classical with a lot of random electromagnetic fields that bring the fluctuations. The theory has its own problems, and may not be the answer, but it gives some idea of what the solution could be, and hopefully quantum phenomena could be explained this way.

\medskip

Roger Balian: I support the opposite viewpoint. Indeed Quantum fluctuations are standard fluctuations, with variance, etc. It's like throwing dices, with the notable exception that this quantum randomness is irreducible because of its underlying non commuting algebra (von Neumann). Nature is random by nature.

\medskip

Marlan O. Scully: I shall give you some examples: Liquid Helium exists in its liquid form because of fluctuations of the atoms. Van der Vaals interactions exist because of vacuum fluctuations (quantized electromagnetic field). The Lamb shift is the real manifestation of this Stochastic Electro Dynamics, as pictured by Boyer and Marshall, but it is important to note that all this vacuum fluctuations can be replaced by radiation reactions, depending on how you write the Hamiltonian. It's the same thing but the physical picture is different. However, explaining Quantum fluctuations with SED is valid only if we consider the subset of problems considered by SED, but if we take another subset of problems, it's no longer the same.

\medskip

Dan C. Cole: I suspect what Marlan is referring to is best illustrated by the work of Peter Milonni, who explored and emphasized the different and complementary roles in QED that is played by vacuum fluctuations and radiation reaction.  Regarding SED, however, I am aware of only one means of working with vacuum fluctuations and radiation reaction.  As for different sets of problems with SED, well, to date SED has been successful with most linear systems of nature, and has had only limited success with the more important category of nonlinear systems in nature.  Hopefully the latter will be resolved, as I discussed in my talk, but of course that remains to be seen.

\medskip

Shahriar S. Afshar : Zero point field and the energy density associated with it are tricky subjects. It is clear that ZPF becomes physically "real", or measurable, when there is radiation reaction. But what about when it is not measured in that sense, when it does not contribute to the physical properties of a test particle? It's just an empty space. The treatment is different, because with radiation reaction I have to treat this energy as real, contributing to the dynamics of the system. Otherwise, without its manifestation as radiation reaction, it cannot be seen as "real", because the energy density would be too high, leading to numerous problems such as a cosmological constant many orders of magnitude lager than the value supported by observations.

\medskip

Theo M. Nieuwenhuizen: The problem of energy density out from Quantum Field theory, together with general relativity, is indeed not understood

\medskip

Marlan O. Scully: This was tried by Puthoff and Sakharov.
 
\medskip 
 
Giacomo Mauro  D Ariano: I would like to draw your attention on the proposal made at this conference by Karl Svozil. A typical manifestation of Quantum fluctuation occurs when an amplification of radiation is made. There is a spontaneous emission that prevents us from using stimulated emission as a cloning process. So, in some way, quantum fluctuation can be seen as a protection from the possibility to increase information by duplication.

\medskip

Luigi Accardi: We should aim at a universal notion of Quantum fluctuations, to pin point the basic difference between classical and quantum physics. In classical physic: there are states of nature in which all observables have no fluctuations. The fundamental difference of Quantum Mechanics is that there exists no such state. On the contrary, in every quantum state, there exist some observables with non zero fluctuations. It is in fact one possible formulation of Heisenberg principle. 

\medskip

Roger Balian: Let me add a point in the same direction. Quantum fluctuations are just a consequence of our inability to describe what nature is made of. We use concepts like position and momentum that are inherited from classical physics. But there are no really such things as position and momentum in nature, it only looks like those properties, so that the mapping does not really fit with what Nature is, and therefore we get fluctuations.

\medskip

Marlan O. Scully: (spoke about single systems, many measurements, noise, ensemble, quantum Langevin)

\medskip

Luigi Accardi: All the Langevin equations are such that when we restrict on the algebra generated by the energy Hamiltonians of the system, we obtain the classical algebra.

\medskip

Marlan O. Scully: No! You obtain non classical properties.

\medskip

Luigi Accardi: Of course, when you consider them on the non classical observables. I'm saying that when you take a non degenerated Hamiltonian, you project directly on the algebra generated by the Hamiltonian. When you consider a larger algebra of observables, of course you have a lot of classical properties. There is a huge quantity of Langevin equations which appear naturally in physics, and they have this property. If you think a posteriori, this is the mathematical explanation of why, at the beginning of quantum theory, all the fundamental physical effects were discovered thinking of classical processes (e.g., Einstein and lasers). The quantum langevin equations were restricted to the energy level of the system, which effectively is classical (Newton).

\medskip

Roger Balian: As Leggett said, there is practically no experimental test of Quantum Mechanics because practically no experiments test non commuting observables. Things are getting different now with entanglement.

\medskip

Al F. Kracklauer: Every charge cannot be isolated from the rest of the universe. This has been so since the Big Bang and presumably will continue until the Big Crunch. Puthoff used this idea to rationalize the SED background. In that context, one might say that Quantum fluctuations are a signature of the equilibrium of all these charges interacting with all the others throughout the universe. The equilibrium part leads to Quantum Mechanics, while the non equilibrium part leads to galaxy formation and all that sort of things.

\medskip

Luis de la Pena: There are basically two schools of thought:
-	For the first one, quantum fluctuations are irreducible, so quantum mechanics gives an exhaustive description of nature. 
-	For the second one, quantum fluctuations can be explained causally. Stochastic electrodynamics is an example of an attempt to explain the phenomena described by Quantum Mechanics causally.
The rest is just details: how we describe, or how we explain.

\medskip

Theo M. Nieuwenhuizen: I think the most important question would be to understand why the hydrogen atom is stable.

\medskip

Andrei Yu. Khrennikov: Before we proceed to the next topic, I would like to hear the point of view of an experimentalist. Gregg, tell us how important are quantum fluctuations for an experimentalist?

\medskip

Gregg Jaeger: Quantum fluctuations are very important. We actually amplify them in our laboratories using Parametric Down Conversion. This process is the main one for the production of entangled quantum states.

\medskip

Satoshi Uchyiama: I know an answer to the question "What is Quantum Fluctuations?" that nobody would contest.

Andrei Yu. Khrennikov: Okay, tell us then.

Satoshi Uchyiama: It's the name of a book by Edward Nelson.

(laughs)

Andrei Yu. Khrennikov: I actually wrote him a few years to invite him to our conferences, but he told me that he doesn't believe anymore in Quantum fluctuations…

\section{CAN QUANTUM MECHANICAL DESCRIPTION BE CONSIDERED COMPLETE?}

The relevance of this vote was questioned before it took place. Prof. Scully 
said for instance that we should actually ask the same question about thermodynamics, 
as a reference test. Prof. Accardi said it was not possible to vote without an agreement 
on a definition of completeness, to which Prof. Khrennikov answered that 
it was clear that everybody has his own. It was decided nevertheless to proceed. 
Khrennikov tried to explicit the question as 

\medskip

{\it "Who believes that Quantum Mechanics is 
the final theory, that there is no deeper theory that would give us a 
deterministic description of reality?"}

\medskip

 at which point Scully  protested that it 
was Khrennikov's own definition of completeness! When asked, Scully said that 
the question should rather be 

\medskip

{\it "Is Quantum Mechanics complete in the 
same sense that Thermodynamics is complete?" }

\medskip

Finally the vote was just proposed as is:

\bigskip

{\it Poll 1: Can Quantum Mechanical description be considered complete?}

-	It is complete :10

-	It is not complete: 19

-	Others : 17

\medskip

During the vote, there was quite a stir and laughs when people noted 
that Marlan Scully had raised his hand twice, both for complete and not complete. Before the second vote took place, Scully explained in what sense completion could be understood for thermodynamics by recalling Einstein's point of view, for whom Thermodynamics was the only subject that was absolutely complete and would never be changed as a body of knowledge within or of itself. Scully remarked that we all know that that thermodynamics has a deeper underlying statistical formulation. Luis de la Peña noted that Einstein was referring to phenomenological thermodynamics, the one that would never change. This is the description that can be considered complete. Scully explained that thermodynamics is a complete body of knowledge, and that it seems natural to say that it won't change as time goes on, but if I one considers quantum thermodynamics, then it looks like  it will indeed change. 

\bigskip

{\bf Poll 2: Can Thermodynamical description be considered complete?}

-	It is complete :12

-	It is not complete: 19

-	Others: 11

\medskip

Surprisingly enough, the same number of votes were obtained for incompleteness of Thermodynamics than for incompleteness of Quantum Mechanics. We had lost four votes in the process.

\medskip

Luis de la Pena: We are speaking about an essential, irreducible incompleteness due to the nature of the description. Of course, every scientific theory is historically incomplete, but this is another kind of incompleteness.

\medskip

Luigi Accardi: Complete doesn't mean final. In two hundred years from now, will Quantum Mechanics be still here?

\medskip

Giacomo Mauro D Ariano: We could indeed make a bet for our grand-grand sons, I would bet that Quantum Mechanics will still be here!

\medskip

(A "young" scientist):  We could perhaps say that a theory is complete when it can describe all known phenomena.

\medskip

Giacomo Mauro D Ariano: It wouldn't work, there would be situations where you could in principle be able to explain a phenomenon with the theory, but you wouldn't be able to do it at all because the calculation would be too complex.

\medskip

Andrei Yu. Khrennikov: I would like to hear Arkady Plotnitsky about the position that Nils Bohr would have adopted in this debate, because it seems quite often to be taken that Bohr thought Quantum Mechanics was complete.

\medskip

Arkady Plotnitsky: Bohr would have voted in the third category, that is, neither complete nor incomplete. He would have said, more rigorously, that Quantum Mechanics is as complete within its scope as classical physics is complete in its scope.

\medskip

Christopher Fuchs: I would support this point of view too, because Quantum Mechanics is, in a certain sense, self contained. So, the question whether or not it is complete doesn't make much more sense than if I would ask whether probability theory is complete or not. I would say that Quantum Mechanics is not going to change in that sense.

\medskip

Marlan O. Scully: I have a comment related to Bohr. In the 1960's, while we were having coffee at night, I asked Gregory Breit,: "Do you think Quantum Mechanics is the be-all and end-all?". And he said that before BCS theory (Bardeen, Cooper, and Schrieffer), he wouldn't have thought so, but after BCS he was overwhelmed and changed his mind. Julian Schwinger experienced the same change. Lamb said that Quantum Mechanics applies only to an ensemble, not to a single system. Furthermore, the wave function does not describe a system, it describes our state of knowledge about that system.

\medskip

Shahriar S. Afshar: Maybe we should qualify this question in the context of how many of us are Bohmians, and how many are adopting a different point of view. For Bohm, all quanta have definite trajectories, and one can indeed do classical (Newtonian) thermodynamics, given the quantum potential etc. are taken into account.

\medskip

Basil J. Hiley: We must be careful here.  For Bohm, all quanta do not have definite trajectories.   Schrödinger and Dirac particles are assumed to follow trajectories.  Photons do not follow trajectories.  Photons must be treated by field theory even in the Bohm approach.

\medskip

Ashok Muthukrishnan: In Quantum Mechanics we have many dualities like Wave/Particle, Unitary Evolution/Collapse, Information/Physical content, and so on. That can be related to fundamental dualities in psychology or philosophy like Freedom/Determinism. The description we have now in terms of the mathematics and of the physical language might be axiomatically complete. Perhaps in the future we should think about merging Quantum mechanics with other domains of knowledge if we don't want to come up in a dead end. It's like the debate between science and religion. A larger language is needed, and it could be complete in that sense.

\medskip

Karl Svozil: I think there is no doubt that there will be a theory that will eventually supersede Quantum Theory. It has to be the case, if not for better reasons then just for historical analogy. There are so many examples throughout history that vividly demonstrate that a theory is never the final answer to everything. For me the real question is more: "Is reality infinitely deep?" To give an image, is reality like an onion,  like Russian doll, where by digging deeper and deeper we will  finally reach something ultimate, or are we bound to endlessly uncover layers after layers without reaching any core? 

\medskip

Andrei Yu. Khrennikov: That's very interesting, but we are moving to a more philosophical ground here, and I must say I have only studied Marxism-Leninism (laughs). I can actually give you the point of view of Lenin on that matter, he said that reality was infinitively deep.

\medskip

Al F. Kracklauer: Kurt Gödel pointed out that the axiom set for arithmetic was likely to be infinite. So, if arithmetic is useful for physics, we might suspect that we would need an infinite set of axioms in physics as well.

\medskip

Yaroslav Volovich: In my opinion, Quantum Mechanics is about quantization. (Thermodynamics is different, it has its own set of problems). Following Newton, we use real numbers and write differential equations to describe physical phenomena and this has proved to be a very useful approach, and similarly the approach which Quantum Mechanics introduces is quantization. It is not impossible to imagine that sometime in the future another such crucial approach will be found that will prove as successful as these two other fundamental approaches. For example, one the of main problems nowadays is with gravitation. In that sense I would say that Quantum Mechanics is probably not the last theory.

\medskip

Giacomo Mauro D Ariano: Suppose Quantum Mechanics is just a syntax or a grammar, basically a set of rules, and that we have the dictionary. It is then possible that we are in front of an onion: we can go deeper and deeper, as discussed by Karl Svozil, but at every layer Quantum Mechanics will hold. We will discover new theories, new particles, but Quantum Mechanics will always remain valid. In the far future, say, in the year 3000, we will still have the same grammar, but with a new dictionary.

\medskip

Shahriar S. Afshar: For me the question would be how many of us actually believe that we will still be able to use the language of wave and particle in the future?

\medskip

Roger Balian: Certainly no theory is final, and Quantum Mechanics will change. However, it won't happen without new phenomena, and if I may be a little bit provocative, I would say that right now it is working so perfectly that it looks like a waste of time to discuss completeness or incompleteness of the theory.

\medskip

Andrei Yu. Khrennikov: I would rather disagree with that. I think on the contrary that we can't wait for new phenomena, precisely because Quantum Mechanics works so perfectly. We need new ideas, otherwise we will test Bell Inequalities for a more hundred years...

\medskip

Giacomo Mauro D Ariano: Maybe I can give an example to illustrate this point. Some of you might have heard of the work of Popescu and Hardy on correlations. They found that there is a whole set of possible theories, in the sense that they yield correlations, that are not causal (and thus offer no possible superluminal communication) but that violate the Cirel'son bound, that is the 2v2 maximum given by Quantum Mechanics. The existence of these super quantum correlations mean that it is actually possible that somebody will find one day something, say new particles, for which the Cirel'son bound is violated.

\medskip

Luigi Accardi: But why would it be so special? There exist uncountably many invariances that are non kolmogorovian and that would provide the same result.

\medskip

Giacomo Mauro D Ariano: That's the point, it means that it is in principle possible that a non kolmogorovian mechanics will supersede Quantum Mechanics.

\medskip

Hans H. Grelland: I believe that increasing our understanding of relations between human beings and of consciousness would improve the understanding of physics in general, and of Quantum Mechanics in particular.

\medskip

At this point, three speakers were given the opportunity to explicit or clarify in short talks some ideas that had been found quite interesting during the conference

Short Talk-1: 
Marlan O. Scully told us how the Maxwell demon paradox was resolved by the Quantum Eraser. He pointed out that it shows a deeper aspect of Quantum Mechanics, namely that Quantum Mechanics is information theoretic, even in a mechanistic sense, and that information is real in a (quantum) physical sense.

\medskip

Short Talk-2:
If the particle has a wave function why wouldn't the wave get a particle function? Pereira told us that, using the HAWKING-ELLIS extended interpretation, the KERR-NEWMAN solution of Einstein's equation can be shown to represent a spinor spacetime structure, whose evolution is governed by the Dirac equation. The KN solution can thus be consistently interpreted as a model for the electron, in which the concepts of mass, charge and spin become linked to the spacetime geometry . In this sense, it can be seen as a concretization of Wheeler's idea of "mass without mass, charge without charge", and also "spin without spin".

\medskip

Short Talk-3: 
Hans H. Grelland showed us his view about the necessity to apply linguistic weak realism in the interpretation of physics (the mathematical formalism of physics, including quantum mechanics, is a proper language).

\medskip

Andrei Yu. Khrennikov: Quantum Mechanics is often said to be a very abstract mathematical theory difficult theory, and that it is one of the features distinguishes it from other theories. However, Quantum Mechanics is just linear algebra. If we consider measure theory, that is, the usual probability theory, it is essentially more abstract, more complex, and possibly deeper than Quantum Mechanics.

\medskip

Roger Balian: Prime numbers are simple, and yet their properties are very complex and amazing at the same time. Still, I agree, the mathematics of probability theory are much more complex than that of Quantum Mechanics.

\medskip

Bob Coecke: Quantum Mechanics is a complex language, even if at the heart it can be rather simple. Consider the language of computers, it's simple on one hand, it's made of 0 and 1. This binary language is also quite complex, impossible to understand, and yet it is at the heart many technological objects that ordinary men can handle.

\medskip

Marlan O. Scully: John Bell said once something like: "wouldn't it be very interesting if all this study on Quantum Mechanics would ultimately lead us to the proof of the existence of God or Buddha?" Have anyone ever come across that quote? I would give 300 hundred dollars for that exact quote…

\medskip

Some participants in the audience had indeed come across that quote, but nobody could pin point exactly where or when John Bell had ventured this daring statement.

\end{document}